\documentclass[epj]{svjour}
\usepackage{graphics}
\begin{document}
\title{Thermodynamics and Quasinormal modes of Park black hole in Ho\v{r}ava gravity}

\author{Jishnu Suresh
\thanks{\emph{email:} jishnusuresh@cusat.ac.in}
\and V C Kuriakose
\thanks{\emph{email:} vck@cusat.ac.in}%
}                     
\offprints{}          
\institute{Department of Physics, Cochin University of Science and Technology, Cochin - 682 022, Kerala, India.}
\date{Received: date / Revised version: date}
%
\abstract{
We study the quasinormal modes of the massless scalar field of Park black hole in the Ho\v{r}ava gravity using the third order 
WKB approximation method and found that black hole is stable against these perturbation. We compare and discuss the results 
with that of Schwarzschild-de Sitter black hole. Thermodynamic properties of Park black hole are investigated and the 
thermodynamic behavior  of upper mass bound is also studied.
} 
\maketitle
\section{Introduction}
\label{intro}
Bardeen, Carter and Hawking formulated four laws of black hole mechanics \cite{Bardeencarter} and Bekenstein introduced the idea of black hole 
entropy \cite{Bekenstein} 
in 1973 and in 1974 Hawking introduced the concept of black hole evaporation \cite{Hawking1} and particle creation by black holes \cite{Hawking2}.
This led to the birth of 
black hole thermodynamics. From the birth itself, it became a source of hope and fascination because it provides a real connection between gravity 
and quantum mechanics. The recent Type Ia supernovae  analysis \cite{Perlmutter} indicated that the expansion of the universe is accelerating and 
the positive cosmological constant could be made responsible for the acceleration of the universe \cite{Caldwell,Garnavich}.
Due to the success of anti-de Sitter (AdS)/conformal field theory (CFT) 
correspondence \cite{Maldacena}
much more attention has been given on studying gravity in the de Sitter (dS) space and asymptotically dS space \cite{Witten}. This leads to an interesting 
proposal which is analogous to the AdS/CFT correspondence in de Sitter space, i.e., dS/CFT correspondence \cite{Strominger1,Strominger2}.
It has been suggested that there is 
a dual relation between quantum gravity on de Sitter (dS) space and Euclidean conformal field theory (CFT) on a boundary of de Sitter space. 

In 2009, Ho\v{r}ava proposed a field theoretic model of gravity as a complete theory in the UV limit \cite{Horava1,Horava2,Horava3}.
This is a a renormalizable theory of gravity in four dimensions. It is non-relativistic in the UV region where as in the IR region
it can be reduced to Einstein's gravity theory with a cosmological constant. Many studies have been done regarding the cosmological
and black hole solutions \cite{LMP,Caicaoohta,KS,Nastase,Kofinas,Calcagni,Park1,Wei,Myung,Myung1,Kim,nv1,nv2,js} of this theory.
In \cite{Park}, Park obtained black  hole and cosmological  solution  by introducing
two parameters $\omega$ and the   cosmological  constant  $\Lambda_W$ and by choosing arbitrary values  for these parameters.
These solutions are analogous to the standard Schwarzschild (A)dS solutions which are absent in the original 
Ho\v{r}ava model. 

Quasinormal modes (QNMs) of a black hole are defined as the solution of the perturbation equations belonging to a certain
complex characteristic frequencies which satisfy the boundary conditions. They govern the decay of perturbations at intermediate times.  
Since they are important when studying the dynamics of black holes they have drawn much more attention in the past years. 
Vishveshwara first put forward the concept of QNMs in the calculations of scattering of gravitational radiation by 
a Schwarzschild black hole \cite{Vishveshwara} and Press \cite{Press} proposed the term quasinormal frequencies. QNM frequencies are the characteristics of a black 
hole since they are independent of initial perturbations and  depend only on black hole parameters, hence one can find information about mass, 
charge and angular momentum \cite{Simone,Konoplya1,Burko,Hod1,Chandrasekhar,Regge}. In addition to this, the properties of QNMs 
have been well explored in the context of 
AdS/CFT correspondence \cite{Horowitz,Wang} and loop quantum gravity \cite{Hod,Dreyer}. As a result of these findings, it is 
widely believed that QNMs carry a unique foot print to directly identify the existence of a black hole. 
For finding the QNMs, we use the third order WKB approximation method \cite{Iyer,Iyerwill}. 
This method was developed by Schutz and Will \cite{Schutz} and later extended to sixth order by Konoplya \cite{Konoplya}.
In this paper we investigate the thermodynamics of Park black holes and also
we investigate the QNMs of massless scalar field around Park black hole and compare with the results of Schwarzschild-de Sitter black hole.

The rest of this paper is organized as follows. In Sect. \ref{sec:3}, we investigate the thermodynamics of Park black hole in  HL gravity and 
Schwarzschild-de Sitter black hole. The comparison of thermodynamics of these black holes and stability are discussed. In
Sect. \ref{sec:4}, the QNMs of Park black hole and SdS black hole are studied and compared the results. Finally, Sect. \ref{sec:5} ends up with a brief 
discussion and conclusion.

\section{Park black hole in Ho\v{r}ava-Lifshitz gravity}
\label{sec:2}

Ho\v{r}ava considered the ADM decomposition of the metric as
\begin{equation}
 ds^2_{4}=-N^2 c^2 dt^2+ g_{ij} \left(dx^i + N^i dt\right)\left(dx^j+ N^j dt\right) ,
 \label{adm}
\end{equation}
where $N$ and $N_{i}$ denote the lapse and shift function, respectively. By introducing an IR modification term $\mu^{4} R^{(3)}$ with an 
arbitrary cosmological constant in Ho\v{r}ava gravity, the modified action can be written as
\begin{eqnarray}
 S &=& \int dt d^{3}x \sqrt{g} N [ \frac{2}{\kappa} \left( K_{ij} K^{ij} -\lambda K^{2} \right)-\frac{\kappa^{2}}{2 \nu^{4}} C_{ij} C^{ij}  \nonumber \\
&+& \frac{\kappa^{2}\mu}{2 \nu^{2}} \epsilon^{ijk} R^{(3)}_{il} \nabla_{j} R^{(3)l}_{k} 
 -\frac{\kappa^2\mu^2}{8} R^{(3)}_{ij} R^{(3)ij}+ \frac{\kappa^2 \mu^2 \omega}{8(3\lambda-1)} R^{(3)} \nonumber \\
 &+& \frac{\kappa^{2} \mu^{2}}{8(3\lambda-1)} \left(\frac{4\lambda-1}{4}(R^{(3)})^2-\Lambda_W R^{(3)}+3 \Lambda_W^{2} \right)
 \label{action}
\end{eqnarray}
where $K_{ij}$ and $C^{ij}$ are the extrinsic curvature and the Cotton tensor, respectively. In the action, 
$\kappa,\nu,\mu,\lambda,\Lambda_{W}, \omega$ are constant parameters. The last term in (\ref{action}) represents a soft violation of 
the detailed balance condition \cite{Horava1}.
For static and spherically symmetric solution, substituting the metric ansatz as 
\begin{equation}
 ds^2=-N(r)^2 c^2 dt^2+\frac{dr^2}{f(r)}+r^2 \left( d\theta^2+ \sin^2 \theta d\phi^2 \right) ,
 \label{line element}
\end{equation}
in the action (\ref{action}) and after angular integration, we obtain the Lagrangian as

\begin{eqnarray}
 {\mathcal{L}} &=& \frac{\kappa^2\mu^2}{8(1-3\lambda)} \frac{N}{\sqrt{f}} [ (2\lambda-1)\frac{(f-1)^2}{r^2}
 -2\lambda \frac{f-1} {r}f' \nonumber \\  
 &+& \frac{\lambda-1}{2}f'^2 - 2 (\omega-\Lambda_W) (1-f-rf') - 3 \Lambda_W^2 r^2 ] .
\label{lagrangain}
\end{eqnarray}
 Kehagias and Sfetsos \cite{KS} obtained only the asymptotically flat solution (with $\Lambda_W=0$) while 
 Mu-In Park \cite{Park} considered an arbitrary $\Lambda_W$ and obtained a general solution. By varying the lapse function 
 $N$ we obtain,
 
 \begin{eqnarray}
(2\lambda-1)&&\frac{(f-1)^2}{r^2} - 2 \lambda \frac{f-1}{r}f' + \frac{\lambda-1}{2} f'^{2} \nonumber \\
&-& 2 (\omega- \Lambda_W) (1-f-rf')- 3 \Lambda_W^{2} r^2 =0 ,
\label{equation of motion1}
 \end{eqnarray}
and similarly by varying $f$, we arrive at
 \begin{eqnarray}
 \left( \frac{N}{\sqrt{f}} \right)' \left( (\lambda-1) f'- 2\lambda \frac{f-1}{r} + 2(\omega-\Lambda_W) r \right) \nonumber \\
 +(\lambda-1) \frac{N}{\sqrt{f}} \left( f''- \frac{2(f-1)}{r^2} \right)=0 .
 \end{eqnarray}
These are the equations of motion.
By giving $\lambda =1$ and solving the field equations, we arrive at the Park solution \cite{Park},
\begin{equation}
 N^2=f_{Park}=1+(\omega-\Lambda_W) r^2-\sqrt{r[\omega (\omega-2 \Lambda_W) r^3 + \beta]} ,
 \label{parksolution}
\end{equation}
where $\beta$ is an integration constant related to the black hole mass. Park's solution can easily be reduced to 
 L\"{u}, Mei, and Pope (LMP)'s solution \cite{LMP} as well as Kehagias and Sfetsos (KS)'s solution \cite{KS}. By choosing
 $\beta= -\alpha^{2} / \Lambda_W$ and $\omega=0$ one can obtain LMP solution as
 \begin{equation}
  f_{LMP}= 1-\Lambda_W r^2- \frac{\alpha}{\sqrt{-\Lambda_W}} \sqrt{r},
  \label{lmpsolution}
 \end{equation}
 And if we choose $\beta= 4 \omega M$ and $\Lambda_W=0$, KS solution is obtained as
 \begin{equation}
  f_{KS}= 1+ \omega r^2- \sqrt{r(\omega^2  r^3 + 4 \omega M)}.
 \end{equation}
Now let us consider the assymptotically dS case of Park solution. In this case, the action is given by an analytic continuation of the action
given by (\ref{action}) \cite{LMP},

\begin{equation}
 \mu \rightarrow i \mu,~\nu^2 \rightarrow -i \nu^2,~\omega \rightarrow -\omega .
 \label{analyticcontinuation}
\end{equation}
From (\ref{parksolution}), for $r \gg [\beta /|\omega ( \omega-2 \Lambda_W)|]^{1/3}$,
i.e., considering assymptotically de Sitter case with $\Lambda_W~\textgreater~0$ and $\omega~\textless~0$, we can arrive at
\begin{equation}
f=1-\frac{\Lambda_W}{2} \left| \frac{\Lambda_W}{ \omega} \right| r^2 - \frac{2 M}{\sqrt{1+2 |\Lambda_W/\omega|}} \frac{1}{r} + {\mathcal O}(r^{-4}) .
\label{dsfofr}
\end{equation}
When we compare (\ref{dsfofr}) with the Schwarzschild-dS solution
\begin{equation}
 f=1-\frac{\Lambda_W}{2} r^2 -\frac{2M}{r} ,
 \label{sds}
\end{equation}
we can see that it agrees up to some numerical factor corrections. In the coming sections we will investigate more about this 
agreement.

\section{Thermodynamics of Park black hole}
\label{sec:3}

\begin{figure}
\resizebox{0.45\textwidth}{!}{%
\includegraphics{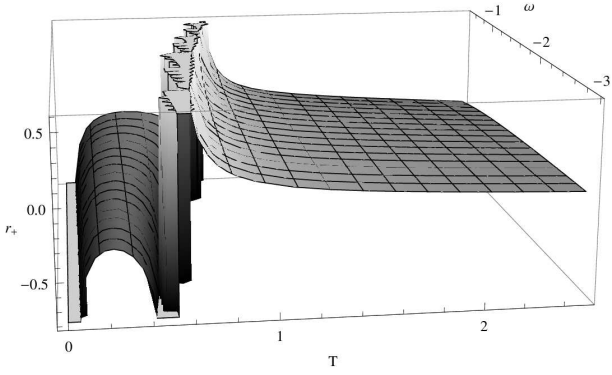}
}
\vspace{1cm}       
\caption{Variation of temperature with horizon radius for different values of $\omega$ with $\Lambda_W=0.0001$.}
\label{figtemp1}       
\end{figure}

In order to explore the  thermodynamics of Park black hole, let us consider (\ref{dsfofr}). In general dS solution 
has two horizons. Larger one $r_{++}$ correspond to the cosmological horizon and the smaller one $r_{+}$ for the black hole horizon. 
By considering the black hole horizon, we can arrive at a relation which connects mass and horizon radius of the black hole,
\begin{equation}
M=\frac{1+ 2 (\omega -\Lambda_W) r_+^2 +\Lambda_W^2 r_{+}^{4}}{4 \omega r_{+}}.
\label{mass1}
\end{equation}
Then from the usual definition of temperature in thermodynamics, we can arrive at temperature of the black hole 
with $\Lambda_W~\textgreater~0$ and $\omega~\textless~0$ as,
\begin{equation}
T=\frac{3 \Lambda_W^{2} r_{+}^{4} + 2(\omega- \Lambda_W) r_{+}^{2} - 1 }{8 \pi r_{+} (1+ (\omega - \Lambda_W) r_{+}^{2})} .
\label{temp1}
\end{equation}
We have plotted the variation of black hole temperature against the horizon radius in fig.\ref{figtemp1}.
From this plot it is evident that there is an infinite  discontinuity in temperature. It occurs at
\begin{equation}
 \tilde{r}_+=\frac{1}{\sqrt{\Lambda_W-\omega}}.
 \label{tempdiscontinuity}
\end{equation}

And for the region $r_{+} \textless \tilde{r}_{+}$, interestingly the temperature is found to be negative. 
The heat capacity of the black hole is given by
\begin{eqnarray}
 C&=&\frac{1}{128 \pi^{3} r_{+}^{4} ((\Lambda_W - \omega) r_{+}^{2}-1)^{3}}  
 [(\Lambda_W r_{+}^{2}-1)^{3} (9 \Lambda_W^{2} r_{+}^{4}-1) \nonumber \\
 &+& 3 r_{+}^{2} (1+ \Lambda_W r_{+}^{2} (8-6\Lambda_W r_{+}^{2}- 3 \Lambda_W^{3} r_{+}^{6})) \omega \nonumber \\
 &-& 12 r_{+}^{4}(1+ \Lambda_W r_{+}^{2}) \omega^{2} + 4 r_{+}^{6} \omega^{3}].
 \label{heatcapacity1}
\end{eqnarray}

\begin{figure}
\resizebox{0.45\textwidth}{!}{
\includegraphics{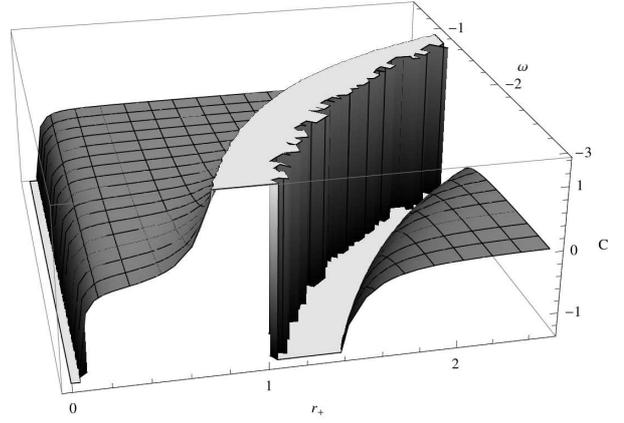}
}
\vspace{1cm}       
\caption{Variation of specific heat with horizon radius for different values of $\omega$ with $\Lambda_W=0.0001$.}
\label{figspec1}      
\end{figure}

In fig.\ref{figspec1}, variation of heat capacity with respect to the black hole horizon $r_{+}$ for different values of coupling
parameters are plotted. From this figure we can see that specific heat undergoes a transition from negative values to positive values or 
in other words black hole changes from a thermodynamically unstable state to a thermodynamically stable state. By looking and comparing the 
two figures, fig.\ref{figtemp1} and fig.\ref{figspec1}, we can straightaway say that in the region where temperature shows the anomalous behaviour due to its 
negative values, the black hole is found to be thermodynamically unstable as its heat capacity is negative.
Now for a Schwarzschild-dS black hole, from (\ref{sds}) we can write
\begin{equation}
  f=1-\frac{\Lambda_W}{2} r^2 -\frac{2M}{r}. \nonumber
\end{equation}
The event horizon is defined by $f(r_{+})=0$,
\begin{equation}
r_{+}-\frac{\Lambda_W}{2} r^3_{+} -2M=0.
\label{horizonsds}
\end{equation}
So we can write the mass as,
\begin{equation}
 M=\frac{r_{+}}{2} - \Lambda_W r_{+}^{3}.
 \label{masssds}
\end{equation}
Using the Bekenstein-Hawking area law,
\begin{equation}
 S=\frac{A}{4}=\pi r_{+}^{2} ,
 \label{bharealaw}
\end{equation}
we can rewrite the mass-horizon relation (\ref{masssds}) as,
\begin{equation}
 M=\frac{1}{2} \sqrt{\frac{S}{\pi}} - \Lambda_W \left(\frac{S}{\pi}\right)^{3/2}.
 \label{massentropysds}
\end{equation}
Using the definition of temperature as $T=\left( \frac{\partial M}{\partial S} \right) $ and that of heat capacity as 
$C = T \left( \frac{\partial S}{\partial T} \right)$, we can arrive at
\begin{equation}
 T=\frac{1}{4 \sqrt{\pi S}}-\frac{3 \Lambda_W \sqrt{S}}{2 \pi^{3/2}},
\end{equation}

\begin{figure}
\resizebox{0.45\textwidth}{!}{
\includegraphics{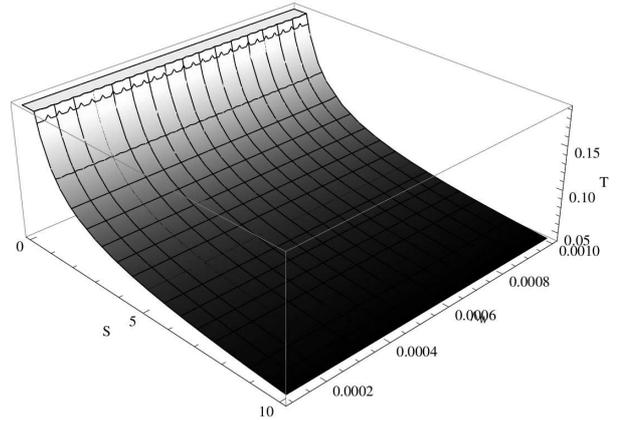}
}
\vspace{1cm}       
\caption{Variation of temperature with entropy for SdS black hole.}
\label{figtemp2}      
\end{figure}

and
\begin{equation}
 C=\frac{\pi^2}{3\Lambda_W \pi+ 18 \Lambda_W ^2 S}-\frac{3 \Lambda_W \sqrt{S}}{2 \pi^{3/2}}-\frac{\pi}{3 \Lambda_W}.
\end{equation}

\begin{figure}
\resizebox{0.45\textwidth}{!}{
\includegraphics{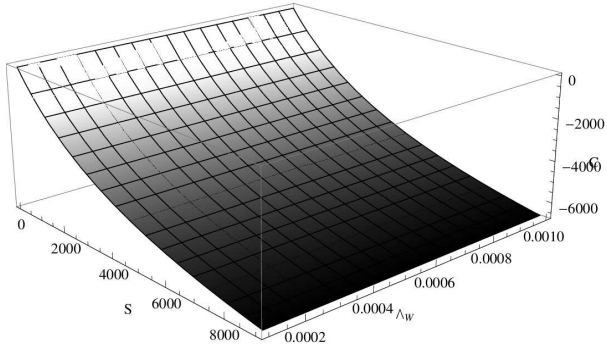}
}
\vspace{1cm}       
\caption{Variation of specific heat with entropy for SdS black hole.}
\label{figspec2}      
\end{figure}

Variation of temperature with respect to entropy is plotted in fig.\ref{figtemp2} while in fig.\ref{figspec2} the variation of heat 
capacity with entropy is plotted. From fig.\ref{figspec2} it is evident that Schwarzschild-dS black hole is thermodynamically 
unstable for all range of entropy values.

Now let us investigate peculiar behavior of Park black hole. As explained in \cite{Park}, for the black hole horizon
to exist and 
curvature singularity at $r=0$ is not naked, one has to consider another condition which is given by,
\begin{equation}
 M \leq \frac{(2 \Lambda_W - \omega)}{4} r_+^3 .
 \label{condition2}
\end{equation}
Or we can say that, $M \textless  \frac{(2 \Lambda_W - \omega)}{4} r_+^3 $ for all $r_{+}$ except for $r_{+}=\tilde{r}_{+}$. And at this point 
mass of the black hole meets the upper bound.

Now let us investigate the thermodynamics of the black hole which has a mass given by the upper mass bound value given in (\ref{condition2})
i.e., 
\begin{equation}
 M_{bound}=\frac{(2 \Lambda_W - \omega)}{4} r_+^3 .
 \label{mass2}
\end{equation}
Using Bekenstein-Hawking area law, we can rewrite (\ref{mass2}) as
\begin{equation}
  M_{bound}=\frac{(2 \Lambda_W - \omega)}{4} \left( \frac{S}{\pi} \right) ^{3/2}.
  \label{massentropyequation}
\end{equation}
From the usual definition of temperature and that of heat capacity, we can arrive at
\begin{equation}
 T=\frac{3(2\Lambda_W -\omega)\sqrt{S}}{8 \pi^{3/2}},
\label{temp2}
 \end{equation}
and
\begin{equation}
 C=2~S.
 \label{heatcapacity2}
\end{equation}
 Since the heat capacity is positive, the black hole having a mass given by (\ref{mass2}) is thermodynamically stable.
 From this fact we can say that the occurrence of infinite discontinuity as well as negative temperature must be due to 
 the existence of a restriction given by (\ref{condition2}) for the mass parameter.
 
\section{Quasinormal modes of Park black hole}
\label{sec:4}

In this part of our work, our main aim is to study the quasinormal modes of massless scalar field of Park black hole in Ho\v{r}ava 
gravity. 
The Klein-Gordon equation for a massless scalar field in this space time is given by
\begin{equation}
 \frac{1}{\sqrt{-g}} \partial_{\mu} \left( \sqrt{-g} g^{\mu \nu} \partial_{\nu} \right) \Phi =0 .
 \label{kleingordon}
\end{equation}
Separating the scalar field in to spherical harmonics, 
\begin{equation}
 \Phi = \frac{1}{r} \Psi(r) Y(\theta,\phi) e^{-i \omega t}
\end{equation}
and using the tortoise coordinate defined by $d r_{*}=\frac{1}{f} dr$, one can obtain the radial equation,
\begin{equation}
 \frac{d^{2} \Psi(r)}{dr_{*}^{2}}+\left[ \omega^{2}- V(r)\right] \Psi(r)=0 ,
 \label{radialequation}
\end{equation}
in which the effective potential $V(r)$ is given by,
\begin{equation}
 V(r)=f(r) \left[ \frac{l(l+1)}{r}+ \frac{1}{r} \frac{\partial f(r)}{\partial r} \right],
 \label{effectivepotential}
\end{equation}
where (\ref{dsfofr}) gives the $f(r)$.

\begin{figure}
\resizebox{0.45\textwidth}{!}{
\includegraphics{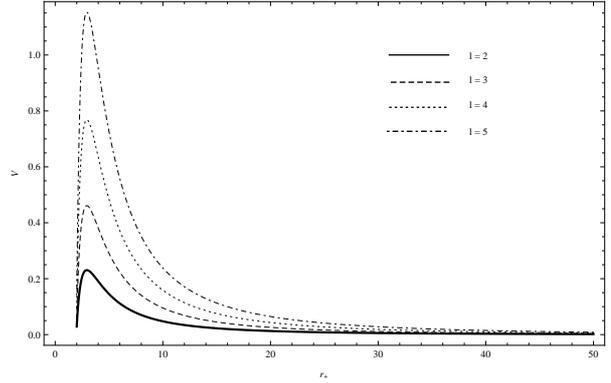}
}
\vspace{1cm}       
\caption{Variation of effective potential with horizon radius for different $l$ values.}
\label{figeffpotential}      
\end{figure}

The behaviour of effective potential for different values of angular quantum number $l$  is plotted in fig.\ref{figeffpotential}. 
From this figure we can say that as the $l$ value increases the potential barrier also get increased. Now we are evaluating 
the quasinormal frequencies of the scalar 
field around Park black hole using the third-order WKB approximation \cite{Schutz,Iyerwill,Iyer}. The formula for 
the complex quasinormal frequencies $\omega_{QNM}$ in this approximation is given by
\begin{equation}
 \omega_{QNM}^{2}=\left[V_{0}+(-2 V_{0}^{''})^{1/2}\Lambda \right] - i(n+\frac{1}{2})(-2 V_{0}^{''})^{1/2} (1+\Omega),
\end{equation}
where
\begin{eqnarray}
 \Lambda&=&\frac{1}{(-2V^{''}_0)^{1/2}} \Bigg\{ \frac{1}{8}\left(\frac{V^{(4)}_0}{V^{''}_0}\right) \left(\frac{1}{4}+\alpha^2\right) \nonumber \\
 &-&\frac{1}{288}\left(\frac{V^{'''}_0}{V^{''}_0}\right)^2 (7+60\alpha^2) \Bigg\},
\end{eqnarray}

\begin{eqnarray}
 \Omega &=&\frac{1}{(-2V^{''}_0)}\Bigg\{\frac{5}{6912}  \left(\frac{V^{'''}_0}{V^{''}_0}\right)^4 (77+188\alpha^2) 
 \frac{1}{384}\left(\frac{V^{'''^2}_0V^{(4)}_0}{V^{''^3}_0}\right) \nonumber \\
 &&(51+100\alpha^2) 
 +\frac{1}{2304}\left(\frac{V^{(4)}_0}{V^{''}_0}\right)^2(67+68\alpha^2) \nonumber \\
  &+&\frac{1}{288} \left(\frac{V^{'''}_0V^{(5)}_0}{V^{''^2}_0}\right)(19+28\alpha^2)-\frac{1}{288} 
   \left(\frac{V^{(6)}_0}{V^{''}_0}\right)(5+4\alpha^2)\Bigg\}
\end{eqnarray}
and 
\begin{eqnarray}
\alpha=n+\frac{1}{2},~~~~~~~~~~~~~~~~~~~~V^{(s)}_0=\frac{d^sV}{dr^s_*}\bigg|_{\;r_*=r_*(r_{p})}
\end{eqnarray}
where $n$ is the overtone and $r_{p}$ is the value of polar coordinate $r$ corresponding to the peak value of effective potential
$V(r)$ given in (\ref{effectivepotential}). Substituting the effective potential (\ref{effectivepotential}) in to the above formula, 
we can obtain the quasinormal frequencies of scalar perturbation for the Park black hole.

\begin{table*}
\centering
\caption{The fundamental $(n=0)$ quasinormal frequencies of scalar filed in the Ho\v{r}ava de Sitter black hole for
$l=1,2,3$ with $\Lambda_W=0.0001$}
\label{table}       
\begin{tabular}{llll}
\hline\noalign{\smallskip}
$\omega$&$\omega_{QNM}(l=1)$&$\omega_{QNM}(l=2)$&$\omega_{QNM}(l=3)$\\
\noalign{\smallskip}\hline\noalign{\smallskip}

-0.5 &0.2916911--0.0981963i& 0.4841694-0.0.096997150i &0.676545-0.09670374i \\

-1.0&0.2914027-0.09809888i&0.4836904-0.09690105i&0.67587629-0.096607976 \\

-1.5&0.2913065-0.09806638i&0.48353071-0.09686899i&0.67565299-0.096576033i \\

-2.0&0.2912584-0.09805013i&0.48345081-0.09685296i&0.675541320-0.096560057i \\

-2.5&0.2912296-0.09804038i&0.48340280-0.09684334i&0.675474301-0.096550470i \\

-3.0&0.2912103-0.09803388i&0.48337089-0.09683693i&0.67542962-0.0965440780i \\

\noalign{\smallskip}\hline
\end{tabular}
\vspace*{0.35cm}  
\end{table*}

\begin{figure}
\resizebox{0.45\textwidth}{!}{
\includegraphics{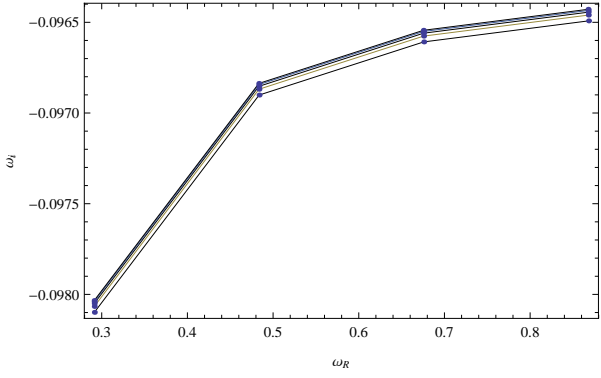}
}
\vspace{1cm}       
\caption{Massless scalar field QNMs of Park black hole for different values of $l$}
\label{figqnm}      
\end{figure}

\begin{figure}
\resizebox{0.45\textwidth}{!}{
\includegraphics{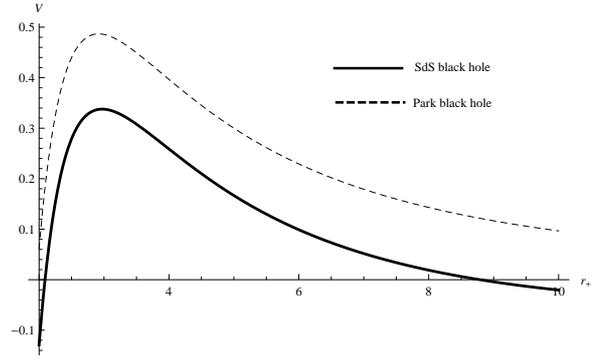}
}
\vspace{1cm}       
\caption{Variation of effective potential of SdS and Park black hole  with horizon radius for a $l=3$.}
\label{figeffpotentialsdshds}      
\end{figure}

The quasinormal frequencies are shown in Table.\ref{table}. Fig.\ref{figqnm} shows that the real and imaginary 
parts of the quasinormal frequencies changes slowly as the the parameter $\omega$ changes for fixed value of $\Lambda_W$.
From Table.\ref{table}, we can find that all frequencies have negative imaginary parts, which means that the black hole 
is stable against these perturbations.

\begin{figure}
\resizebox{0.45\textwidth}{!}{
\includegraphics{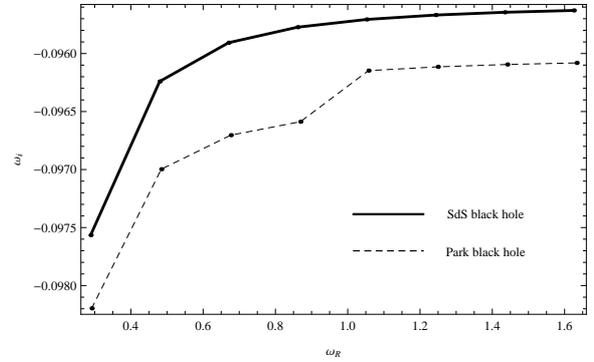}
}
\vspace{1cm}       
\caption{Massless scalar field QNMs of SdS and Park black holes for $l=3$.}
\label{figqnmcompare}      
\end{figure}

Let us now consider the Schwarzchild-de Sitter case. In fig.\ref{figeffpotentialsdshds} we have plotted the effective potential for 
the Schwarzchild-de Sitter case along with the Park black hole case for a particular $l$ value. And from this figure we
can see that the behaviour is almost the same. Applying the third-order WKB 
approximation method to evaluate the fundamental quasinormal modes and from the obtained quasinormal 
mode frequencies of SdS black hole and Park black hole, we can compare them. In fig.\ref{figqnmcompare} we have plotted the fundamental
quasinormal mode frequencies for fixed values of $\Lambda_W$ and $\omega$ for these two black holes. 
From this it is evident that, both black holes show similar behaviour against 
these massless scalar perturbations. So as in the case of Park black hole, SdS black hole are also stable against these perturbations.

\section{Discussion and conclusion}
\label{sec:5}
In this paper we have investigated the quasinormal modes of the massless scalar field of Park black hole. QNMs are
studied using the third-order WKB approximation method. We have found that the frequencies all have negative imaginary parts. 
Hence the Park black hole is stable against these perturbations. These results are compared with those of
Schwarzschild-de Sitter case and find that they have almost the same behavior. We have also studied the thermodynamic aspects of
Park black hole and find that it shows negative values as well as infinite discontinuity in temperature. This may be due to the 
existence of an upper bound of the mass parameter. In addition to this we have studied the thermodynamic properties of Park black hole which
has a mass equal to the upper bound value of the mass parameter and found that these black holes are 
thermodynamically stable. When we study the perturbative effects on Park and SdS black hole, there is an agreement between them up to some numerical factor 
corrections, while they have entirely different thermodynamic
behaviors.

\section*{Acknowledgements}
The authors wish to thank UGC, New Delhi for financial
support through a major research project sanctioned to VCK. VCK also
wishes to acknowledge Associateship of IUCAA, Pune, India.

%
%
%

\begin{thebibliography}{}

\bibitem{Bardeencarter} Bardeen J. M., Carter. B., Hawking. S. W., Commun. Math. Phys., \textbf{31} (1973) 161.
\bibitem{Bekenstein} Bekenstein J. D., Phys. Rev. D., \textbf{7} (1973) 2333.
\bibitem{Hawking1} Hawking S. W., Nature., \textbf{248} (1974) 30.
\bibitem{Hawking2} Hawking S. W., Commun. Math. Phys., \textbf{43} (1975) 43.
\bibitem{Perlmutter} Perlmutter. S., et al., Astrophys. J., \textbf{483} (1997) 565.
\bibitem{Caldwell} Caldwell. R. R., Dave. R., Steinhardt. P. J., Phys. Rev. Lett., \textbf{80} (1998) 1582.
\bibitem{Garnavich} Garnavich. P. M., et al., Astrophys. J., \textbf{509} (1998) 74.
\bibitem{Maldacena} Maldacena. J. M., Adv. Theor. Math. Phys., \textbf{2} (1998) 231
\bibitem{Witten} Witten. E., hep-th/0106109.
\bibitem{Strominger1}  Strominger. A., JHEP., \textbf{0111} (2001) 0341.
\bibitem{Strominger2} Strominger2. A., JHEP., \textbf{0111} (2001) 0491.
\bibitem{Horava1} Ho\v{r}ava P., Phys. Rev. D, \textbf{79} (2009) 084008.
\bibitem{Horava2} Ho\v{r}ava P., JHEP, \textbf{03} (2009) 020.
\bibitem{Horava3} Ho\v{r}ava P., Phys. Rev. Lett., \textbf{102} (2009) 161301.
\bibitem{LMP} Lu H., Mei  J, Pope C. N., Phys. Rev. Lett., \textbf{103} (2009) 091301.
\bibitem{Caicaoohta} Cai R. G., Cao L. M, Ohta N., Phys. Rev. D, \textbf{81} (2009) 024003.
\bibitem{KS} Kehagias A, Sfetsos K., Phys. Lett. B, \textbf{678} (2009) 123.
\bibitem{Nastase} Nastase. H., Phys. Lett. B, \textbf{67} (2009) 123.
\bibitem{Kofinas} Kiritsis E, Kofinas G., Nucl. Phys. B, \textbf{821} (2009) 467.
\bibitem{Calcagni} Calcagni G., JHEP, \textbf{09} (2009) 112.
\bibitem{Park1} Park M. i., JCAP, \textbf{01} (2010) 001.
\bibitem{Wei} Wei S. W, Liu Y. X, Guo H., EPL, \textbf{99} (2012) 20004.
\bibitem{Myung} Myung Y. S., Phys. Lett. B. \textbf{684} (2010) 158.
\bibitem{Myung1} Myung Y. S., Phys. Lett. B. \textbf{678} (2009) 127.
\bibitem{Kim} Myung Y. S, Kim Y. W., Eur. Phys. J. C. \textbf{68} (2010) 265.
\bibitem{nv1} Nijo Varghese, V. C. Kuriakose., Mod. Phys. Lett. A, \textbf{26} (2011) 1645.
\bibitem{nv2} Nijo Varghese, V. C. Kuriakose., Gen. Relativ. Gravit., \textbf{43} (2011) 2755.
\bibitem{js} Jishnu Suresh, V. C. Kuriakose., Gen. Relativ. Gravit., \textbf{45} (2013) 1877.
\bibitem{Park} Park M. i., JHEP, \textbf{0909} (2009) 123.
\bibitem{Vishveshwara} Vishveshwara. C. V., Nature., \textbf{227} (1970) 936.
\bibitem{Press} Press. W. H., Astrophys. J., \textbf{170} (1971) 105.
\bibitem{Simone} Simone. L. E., Will. C. M., Classical Quant. Grav., \textbf{9} (1992) 963.
\bibitem{Konoplya1} Konoplya. R. A., Phys. Lett. B., \textbf{550} (2002) 117.
\bibitem{Burko} Burko. R. A., Khanna. G., Phys. Rev. D., \textbf{70} (2004) 044018.
\bibitem{Hod1} Hod. S., Phys. Rev. D., \textbf{84} (2011) 044046.
\bibitem{Chandrasekhar} Chandrasekhar. S., Detweiler. S. L., P. Roy. Soc. A-Math. Phys., \textbf{344} (1975) 441.
\bibitem{Regge} Regge. T., Wheeler. J. A., Phys. Rev., \textbf{108} (1999) 1063.
\bibitem{Horowitz} Horowitz. G. T., Hubeny. V. E., Phys. Rev. D., \textbf{62} (2000) 024027.
\bibitem{Wang} Wang. B., et al., Phys. Rev. D., \textbf{70} (2004) 064025.
\bibitem{Hod} Hod. S., Phys. Rev. Lett., \textbf{81} (1998) 4293.
\bibitem{Dreyer} Dreyer. O., Phys. Rev. Lett., \textbf{90} (2003) 081301.
\bibitem{Iyer} Iyer. S., Phys. Rev. D., \textbf{35} (1987) 3632.
\bibitem{Iyerwill} Iyer. S, Will. C. M., Phys. Rev. D., \textbf{35} (1987) 3621.
\bibitem{Schutz} Schutz. B. F, Will. C. M., Astrophys. J. Lett. Ed. \textbf{291} (1985) 33.
\bibitem{Konoplya} Konoplya. R. A., Phys. Rev. D., \textbf{68} (2003) 024018.

\end{thebibliography}
%

\end{document}